\documentclass[usenatbib]{mn2e} \usepackage{psfig}

\title[ULTRACAM photometry of XZ~Eri and DV~UMa] {ULTRACAM photometry of the
  eclipsing cataclysmic variables XZ~Eri and DV~UMa}

\author[W.\,J.\,Feline et al.] {W.\,J.\ Feline,$^1$\thanks{E-mail:
  w.feline@shef.ac.uk} V.\,S.\ Dhillon,$^1$ T.\,R.\ Marsh$^{2}$ and
  C.\,S.\ Brinkworth$^{3}$\\ $^1$Department of Physics and Astronomy,
  University of Sheffield, Sheffield, S3 7RH, UK\\ $^2$Department of
  Physics, University of Warwick, Coventry CV4 7AL, UK\\
  $^3$Department of Physics and Astronomy, University of Southampton,
  Southampton, SO17 1BJ, UK\\}

\date{\center{\Large Accepted for publication in the Monthly
    Notices of the Royal Astronomical Society}}

\begin{document}
  \maketitle
  
\begin{abstract}
We present high-speed, three-colour photometry of the faint eclipsing
cataclysmic variables XZ~Eri and DV~UMa. We determine the system
parameters through two techniques: first, timings of the eclipse
contact phases of the white dwarf and bright-spot using the derivative
of the light curve; and secondly, a parameterized model of the eclipse
fitted to the observed light curve by $\chi^{2}$ minimisation. For
both objects, we prefer the latter method, as it is less affected by
photon noise and rapid flickering. For XZ~Eri we obtain a mass ratio
$q=0.1098\pm0.0017$ and an orbital inclination
$i=80\fdg16\pm0\fdg09$. For DV~UMa we derive figures of
$q=0.1506\pm0.0009$ and $i=84\fdg24\pm0\fdg07$. The secondary star in
XZ~Eri has a very low mass $M_{\rmn{r}}/M_{\sun}=0.0842\pm0.0024$,
placing it close to the upper limit on the mass of a brown dwarf.
\end{abstract}


\section{Introduction}
\label{introduction}

Cataclysmic variable stars (CVs) are a class of interacting binary
system undergoing mass transfer via a gas stream and accretion disc
from a Roche-lobe filling secondary to a white dwarf primary. A bright
spot is formed at the intersection of the disc and gas stream, giving
rise to an `orbital hump' in the light curve at phases $0.6-1.0$ due
to foreshortening of the bright-spot. The light curves of eclipsing
CVs can be quite complex, with the accretion disc, white dwarf and
bright-spot all being eclipsed in rapid succession. With sufficient
time-resolution, however, this eclipse structure allows the system
parameters to be determined to a high degree of
accuracy. \citet{warner95} gives a comprehensive review of CVs.

The class of CVs known as the dwarf novae intermittently undergo
outbursts -- increases in brightness of between 2--5 magnitudes. Both
XZ~Eri and DV~UMa are members of the SU~UMa sub-class of dwarf novae,
which also exhibit superoutbursts (about 0.7 mag brighter than normal
outbursts) at semiregular intervals.

XZ~Eri was first noted to be variable by \citet{shapley34}. Until
recently \citep{howell91,szkody92}, however, XZ~Eri had been rather
poorly studied. The presence of eclipses in the lightcurve of XZ~Eri
was discovered by \citet{woudt01}. More recently, \citet{uemura04}
observed superhumps in the outburst lightcurve of XZ~Eri, confirming
its classification as an SU~UMa star.

Previous observations of DV~UMa are summarized by \citet{nogami01},
who also present light curves obtained during the 1995 outburst and
the 1997 superoutburst. \citet{patterson00} present superoutburst and
quiescent photometry from which they derive the system
parameters. \citet{mukai90} estimated the spectral type of the
secondary star to be $\sim$M4.5 from spectroscopic observations.

In this paper we present simultaneous three-colour, high-speed
photometry of XZ~Eri and DV~UMa. We derive the system parameters via
two separate methods -- timings of the eclipse contact phases and
fitting a parameterized model of the eclipse -- and discuss the
relative merits of each.

%__________________________________________________________________

\section{Observations}
\label{observations}

\begin{figure}
\centerline{\psfig{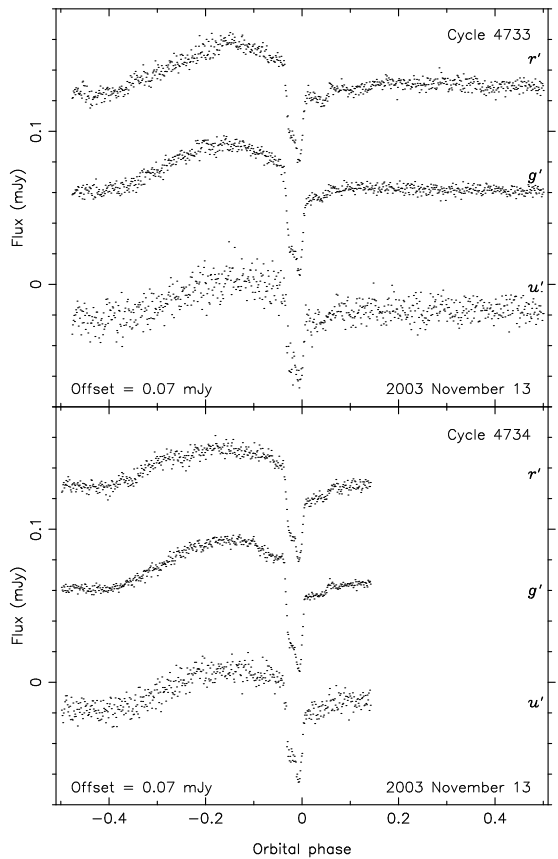}}
\caption{The light curve of XZ~Eri. The data are contiguous. The {\em
  r}$^{\prime}$ data are offset vertically upwards and the {\em
  u}$^{\prime}$ data are offset vertically downwards.}
\label{light curve}
\end{figure}

\begin{figure}
\centerline{\psfig{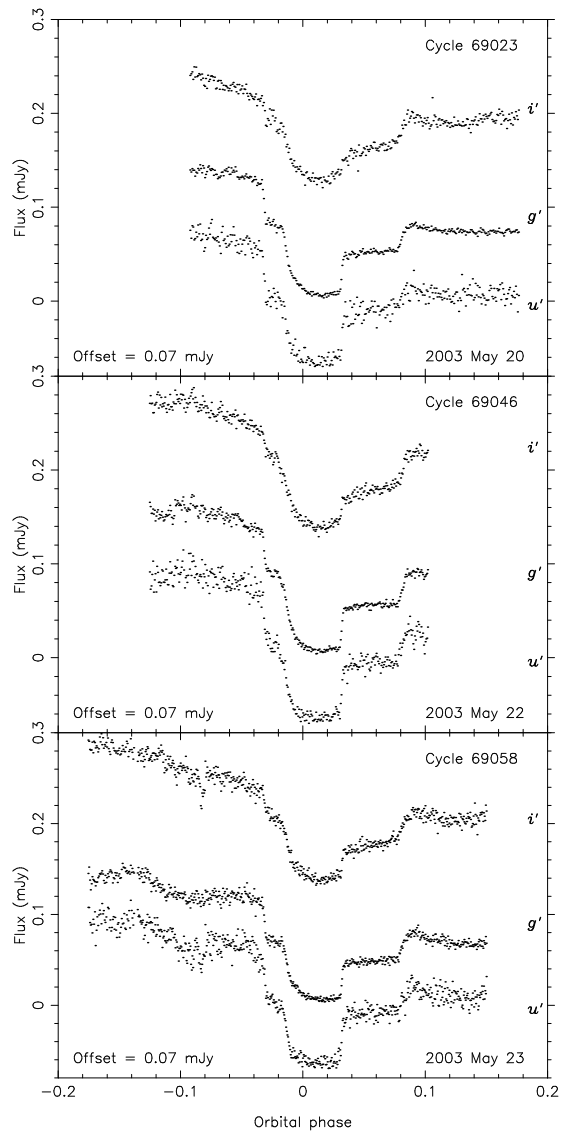}}
\caption{The light curve of DV~UMa. The {\em  i}$^{\prime}$ data are
  offset vertically upwards and the {\em  u}$^{\prime}$ data are
  offset vertically downwards.}
\label{dvuma_light curve}
\end{figure}

XZ~Eri and DV~UMa were observed simultaneously in three colour bands
using ULTRACAM (\citealt{dhillon01b}; Dhillon et~al., in preparation)
on the 4.2-m William
Herschel Telescope (WHT) at the Isaac Newton Group of Telescopes, La
Palma. The observations are summarized in Table~\ref{journal}. Data
reduction was carried out as described in \citet{feline04a} using the
ULTRACAM pipeline data reduction software. The resulting light curves
of XZ~Eri and DV~UMa are shown in Figs~\ref{light curve} and
\ref{dvuma_light curve}, respectively. The observations of XZ~Eri
began at high airmass (1.8) -- this is evident in the improved quality
of the second cycle. Note also that the XZ~Eri data of 2003 November
13 have significantly worse time-resolution than those of DV~UMa,
despite both objects being of similar magnitude. This is due to the
higher brightness of the sky on 2003 November 13.

\section{Light-curve morphology}
\label{morphology}

The light curve of XZ~Eri shown in Fig.~\ref{light curve} is a
classic example of an eclipsing dwarf nova. Between phase --0.4 and
the start of eclipse, the orbital hump is clearly visible, with a
brightening in {\em g}$^{\prime}$ flux of 0.025 mJy (0.5 mag). The
light curve clearly shows separate eclipses of the white dwarf and
bright-spot in all three colour bands.

During our observations XZ~Eri had {\em g}$^{\prime}\sim19.5$, falling
to {\em g}$^{\prime}\sim21.5$ in mid-eclipse. Comparing this to the
previous (quiescent) observations of \citet{woudt01}, who observed the
system at $\rmn{V}\sim19.2$, confirms that XZ~Eri was in quiescence at the
time of our observations.

The light curve of DV~UMa is presented in Fig.~\ref{dvuma_light
curve}. Although the phase coverage is less complete than for XZ~Eri,
the eclipse morphology is again typical of eclipsing short-period
dwarf novae. The white dwarf and bright-spot ingress and egress are
both clear and distinct. The orbital hump in DV~UMa is much less
pronounced than in XZ~Eri.

\citet{howell88} quoted $\rmn{V}\sim19.2$ in quiescence for DV~UMa. This
compares to {\em g}$^{\prime}\sim19$ at the time of our observations,
which fell to {\em g}$^{\prime}\sim22$ during eclipse. DV~UMa was
therefore in quiescence over the course of our observations.

\begin{table*}
\begin{center}
\caption[]{Journal of observations. Observing conditions were clear
  except for 2003 May 20, when thin cirrus was present. The dead-time
  between exposures was 0.025~s for all the DV~UMa observations, and
  0.024~s for the XZ~Eri observations.}
\begin{tabular}{ccccccccccc}
\hline
Date & Cycle & Target & Filters & Exposure time (s) & 
Data points & Eclipses & Seeing (arcsec) & Airmass \\
\hline
2003 May 20 & 69023 & DV~UMa & {\em u}$^{\prime}${\em g}$^{\prime}${\em
  i}$^{\prime}$ & 5.921 & 339 & 1 & 1.3--2.0 & 1.5--1.8 \\
2003 May 22 & 69046 & DV~UMa & {\em u}$^{\prime}${\em g}$^{\prime}${\em
  i}$^{\prime}$ & 4.921 & 345 & 1 & 1.2 & 1.4--1.5 \\
2003 May 23 & 69058 & DV~UMa & {\em u}$^{\prime}${\em g}$^{\prime}${\em
  i}$^{\prime}$ & 4.921 & 60 & 0 & 1.0 & 1.6--1.9 \\
2003 May 23 & 69058 & DV~UMa & {\em u}$^{\prime}${\em g}$^{\prime}${\em
  i}$^{\prime}$ & 3.921 & 540 & 1\\
2003 Nov.\ 13 & 4733/4734 & XZ~Eri & {\em u}$^{\prime}${\em g}$^{\prime}${\em
  r}$^{\prime}$ & 6.997 & 1225 & 2 & 1.0--2.0 & 1.4--1.8 \\
\hline
\end{tabular}
\label{journal}
\end{center}
\end{table*}

\section{Orbital ephemerides}
\label{ephemeris}

The times of white dwarf mid-ingress $T_{\rmn{wi}}$ and mid-egress
$T_{\rmn{we}}$ were determined by locating the times when the minimum
and maximum values, respectively, of the light curve derivative
occurred \citep*{wood85}. The times of mid-eclipse $T_{\rmn{mid}}$
given in Table~\ref{eclipse_times} were determined by assuming the
white dwarf eclipse to be symmetric around phase zero and taking
$T_{\rmn{mid}}=(T_{\rmn{we}}+T_{\rmn{wi}})/2$.

To determine the orbital ephemeris of XZ~Eri we used the one
mid-eclipse time of \citet{woudt01}, the 25 eclipse timings of
\citet{uemura04} and the six times of mid-eclipse given in
Table~\ref{eclipse_times}. We used errors of $5\times10^{-4}$~d for
the \citet{woudt01} data, $1\times10^{-4}$~d for the
\citet{uemura04} timings and $4\times10^{-5}$~d for the ULTRACAM
data. A linear least-squares fit to these times gives:

\begin{table}
\begin{center}
\caption[]{Mid-eclipse timings (HJD + 2452000).}
\begin{tabular}{cccc}
\hline
XZ~Eri cycle & {\em u}$^{\prime}$ & {\em g}$^{\prime}$ & {\em r}$^{\prime}$\\
\hline
4733 & 957.508910 & 957.508789 & 957.508870\\
4734 & 957.570081 & 957.570000 & 957.570000\\
\hline
DV~UMa cycle & {\em u}$^{\prime}$ & {\em g}$^{\prime}$ & {\em i}$^{\prime}$\\
\hline
69023 & 780.469225 & 780.469225 & 780.469225\\
69046 & 782.443801 & 782.443829 & 782.443801\\
69058 & 783.474062 & 783.474040 & 783.474040\\
\hline
\end{tabular}
\label{eclipse_times}
\end{center}
\end{table}

\begin{displaymath}
\begin{array}{ccrcrl}
\\ HJD & = & 2452668.04099 & + & 0.061159491 & E.  \\
 & & 2 & \pm & 5 &
\end{array} 
\end{displaymath}

The orbital ephemeris of DV~UMa was determined in a similar way using
the 18 mid-eclipse timings of \citet{nogami01}, the 12 timings of
\citet{howell88}, the 12 timings of \citet{patterson00} and the nine
times of mid-eclipse given in Table \ref{eclipse_times}, with errors
of $5\times10^{-4}$~d assigned to the data of \citet{nogami01} and
\citet{howell88}, $1\times10^{-4}$~d to the data of
\citet{patterson00} and $4\times10^{-5}$~d to the ULTRACAM data. A
linear least-squares fit to these times gives:

\begin{displaymath}
\begin{array}{ccrcrl}
\\ HJD & = & 2446854.66157 & + & 0.0858526521 & E.
\\ & & 9 & \pm & 14 &
\end{array} 
\end{displaymath}

These ephemerides were used to phase all of our data.

\begin{figure*}
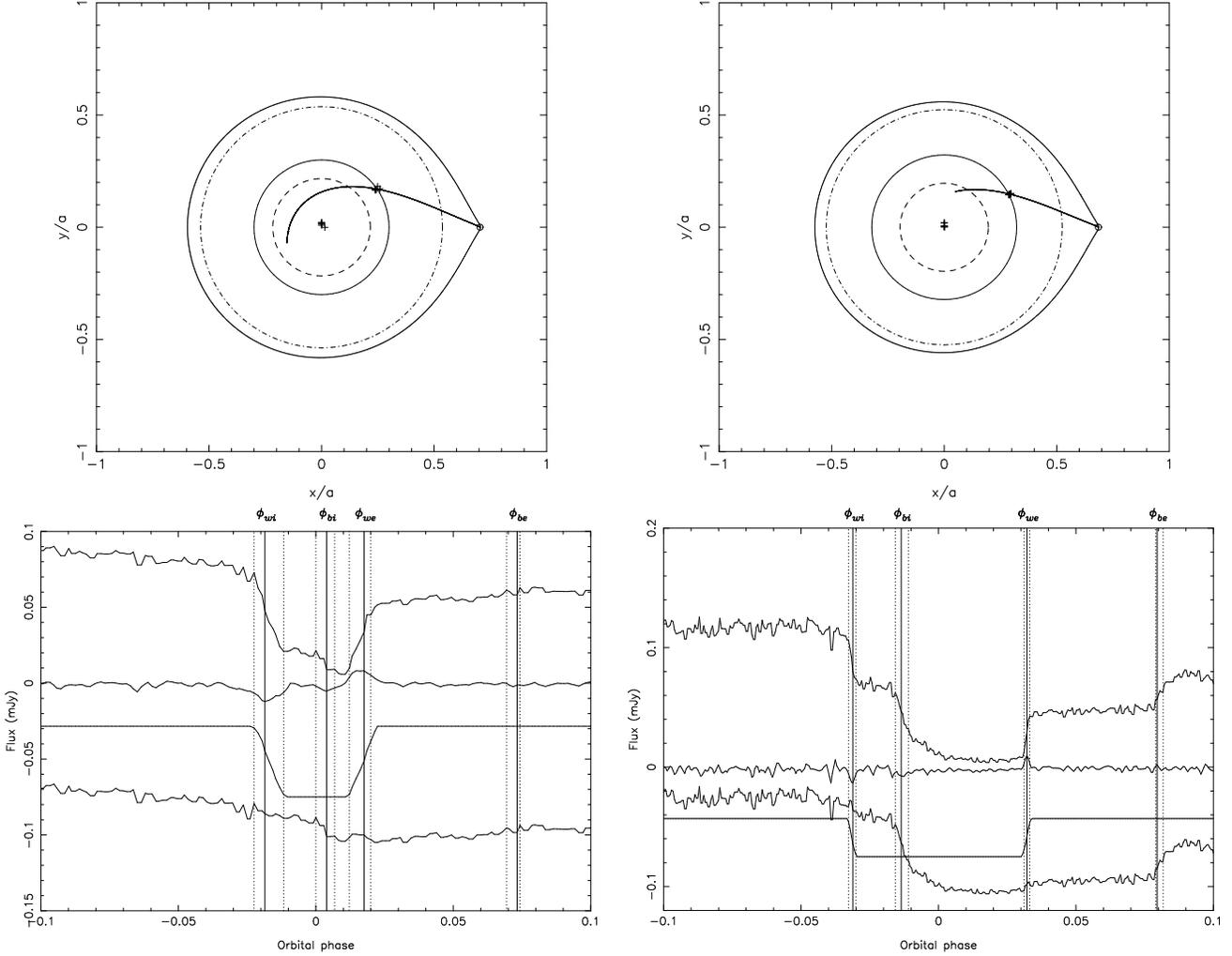

\begin{tabular}{cc}
\psfig{figure=figure3a.ps,width=7.0cm,angle=-90.} &
\psfig{figure=figure3b.ps,width=7.0cm,angle=-90.}
\\
\psfig{figure=figure3c.ps,width=8.4cm,angle=-90.} &
\psfig{figure=figure3d.ps,width=8.4cm,angle=-90.}
\end{tabular}
\caption{Top row: Trajectory of the gas stream from the secondary star
  for (left) XZ~Eri ($q=0.117$, $i=80\fdg3$, $R_{\rmn{d}}/a=0.300$ and
  $R_{\rmn{circ}}/a=0.217$) and (right) for DV~UMa ($q=0.148$,
  $i=84\fdg4$, $R_{\rmn{d}}/a=0.322$ and
  $R_{\rmn{circ}}/a=0.196$). The Roche lobe of the primary, the position
  of the inner Lagrangian point $\rmn{L}_{1}$ and the disc of radius
  $R_{\rmn{d}}$ are all plotted. The positions of the white dwarf and
  bright-spot light centres corresponding to the observed ingress and
  egress phases are also plotted. The circularisation radius
  $R_{\rmn{circ}}$ \citep[][ equation 13]{verbunt88} is shown as a
  dashed circle, and the tidal radius \citep{paczynski77} as a
  dot-dashed circle. Bottom row: White dwarf deconvolution of (left) the
  {\em g}$^{\prime}$ band light curve of XZ~Eri on 2003 November 13
  (cycle 4733) and (right) the {\em g}$^{\prime}$ band light curve of
  DV~UMa on 2003 May 23. Top to bottom: The data after smoothing by a
  median filter; the derivative after smoothing by a box-car filter
  and subtraction of a spline fit to this, multiplied by a factor of
  1.5 for clarity; the reconstructed white dwarf light curve, shifted
  downwards by 0.075 mJy; the original light curve minus the white
  dwarf light curve after smoothing by a median filter, shifted
  downwards by 0.11 mJy. The vertical lines show the contact phases of
  the white dwarf and bright-spot eclipses, the dotted lines
  corresponding to $\phi_{\rmn{w}1},\ldots,\phi_{\rmn{w}4}$,
  $\phi_{\rmn{b}1},\ldots,\phi_{\rmn{b}4}$ and the solid lines
  (labelled) to $\phi_{\rmn{wi}}$, $\phi_{\rmn{we}}$ and
  $\phi_{\rmn{bi}}$, $\phi_{\rmn{be}}$. The bright-spot ingress and
  egress are plainly visible in the light curves of both objects,
  quickly following the white dwarf ingress and egress, respectively.}
\label{timings}
\end{figure*}

\section{Light-curve decomposition}
\label{decomposition}

\subsection{The derivative method}
\label{techniques_times}
This method of determining the system parameters of an eclipsing dwarf
nova was originally developed by \citet{wood86b}. It relies upon the
fact that there is a unique relationship between the mass ratio
$q=M_{\rmn{r}}/M_{\rmn{w}}$ and orbital inclination {\em i} for a
given eclipse phase width $\Delta\phi$ \citep{bailey79}.

The eclipse contact phases given in Tables~\ref{wd_times} and
\ref{bs_times} were determined using the derivative of the light
curve, as described by \citet[][and references therein]{feline04a}. The
midpoints of ingress and egress are denoted by $\phi_{\rmn{i}}$ and
$\phi_{\rmn{e}}$, respectively. The eclipse contact phases
corresponding to the start and end of the ingress are denoted
$\phi_{1}$ and $\phi_{2}$, and the start and end of the egress by
$\phi_{3}$ and $\phi_{4}$. In the discussion that follows, we use the
suffixes `$\rmn{w}$' and `$\rmn{b}$' to denote white dwarf and
bright-spot contact phases, respectively (e.g.\ $\phi_{\rmn{wi}}$
means the mid-point of the white dwarf ingress). The eclipse phase
full width at half-depth is $\Delta\phi = \phi_{\rmn{we}} -
\phi_{\rmn{wi}}$. In the following analysis we have combined the
timings of all three colour bands for each target in order to increase
the accuracy of our results.

The mass ratio -- and hence the inclination -- may be determined by
comparing the bright-spot light centres corresponding to the measured
eclipse contact phases $\phi_{\rmn{bi}}$ and $\phi_{\rmn{be}}$ with
the theoretical stream trajectories for different mass ratios. As
illustrated in Fig.~\ref{timings}, the assumption that the gas stream
(originating from the inner Lagrangian point $\rmn{L}_{1}$) passes directly
through the light centre of the bright-spot at the edge of the disc
allows the determination of the mass ratio, orbital inclination and
relative outer disc radius $R_{\rmn{d}}/a$ (where $a$ is the
orbital separation).

For the mean eclipse phase width of $\Delta\phi = 0.035889$, the
eclipse timings of XZ~Eri (Tables~\ref{wd_times} and \ref{bs_times})
yield the mass ratio, inclination and relative disc radius given in
Table~\ref{parameters}. The results for DV~UMa for the mean eclipse
phase width of $\Delta\phi = 0.063604$ are also given in
Table~\ref{parameters}.  The errors on these parameters are determined
by the rms variations in the measured contact phases. We use the
bright-spot eclipse timings to determine upper limits on the angular
size and the radial and vertical extent of the bright-spots, defining
$\Delta \theta$, $\Delta R_{\rmn{d}}$, $\Delta Z$ and $\Delta Z_{2}$
as in \citet{feline04a}. The mean position and extent of the bright
spots thus derived are given in Table~\ref{bs}.

Using the mass ratio and orbital inclination given in
Table~\ref{parameters} and the eclipse constraints on the radius of
the white dwarf (Table~\ref{wd_times}), we find that the white dwarf
in XZ~Eri has a radius of $R_{\rmn{w}}/a=0.012 \pm 0.002$. For DV~UMa
we obtain $R_{\rmn{w}}/a=0.0075 \pm 0.0020$. We will continue
under the assumption that the eclipsed central object is a bare white
dwarf. This assumption and its consequences are discussed in more
detail in \citet{feline04a}.

The fluxes given in Table~\ref{wd_times} were fitted to the
hydrogen-rich, $\log g=8$ white dwarf model atmospheres of
\citet*{bergeron95}. The colour indices quoted therein were converted
to the SDSS system using the observed transformations of
\citet{smith02}. The white dwarf temperatures $T_{\rmn{w}}$ thus
calculated are given in Table~\ref{parameters}.

To determine the remaining system parameters of XZ~Eri and DV~UMa we
have used the Nauenberg mass--radius relation for a cold, non-rotating
white dwarf \citep{nauenberg72, cook84}, which gives an analytical
approximation to the Hamada--Salpeter mass--radius relation
\citep{hamada61}. This relation, together with Kepler's third law and
the relative white dwarf radius, allows the analytical determination
of the absolute system parameters, given in
Table~\ref{parameters}. The secondary radius $R_{\rmn{r}}$ has been
calculated by approximating it to the volume radius of the Roche lobe
\citep{eggleton83}, which is accurate to better than 1 per~cent.
Because the Nauenberg mass-radius relation assumes a cold white dwarf,
we have attempted to correct this to a temperature of $T_{\rmn{w}}\sim
15000$~K for XZ~Eri and to $T_{\rmn{w}}\sim 20000$~K for DV~UMa, the
approximate temperatures given by the model atmosphere fit. The radius
of the white dwarf at $10000$~K is about 5 per~cent larger than for a
cold ($0$~K) white dwarf \citep{koester86}. To correct from $10000$~K
to the appropriate temperature, the white dwarf cooling curves of
\citet{wood95} for $M_{\rmn{w}}/M_{\sun}=1.0$, the approximate masses
given by the Nauenberg relation, were used. This gave total radial
corrections of $6.0$ and $7.0$ per~cent for XZ~Eri and DV~UMa,
respectively.

\begin{table*}
\begin{center}
\caption[]{White dwarf contact phases and out-of-eclipse white dwarf
  fluxes. We estimate that the errors on the fluxes are $\pm0.001$mJy.}
\begin{tabular}{ccccccccc}
\hline
Cycle & Band & $\phi_{\rmn{w}1}$ & $\phi_{\rmn{w}2}$ &
$\phi_{\rmn{w}3}$ & $\phi_{\rmn{w}4}$ &
$\phi_{\rmn{wi}}$ & $\phi_{\rmn{we}}$ & Flux (mJy)\\
\hline
XZ~Eri\\
4733 & {\em u}$^{\prime}$ & --0.020996 & --0.015625 & 0.016113
 & 0.022461 & --0.018555 & 0.018555 & 0.0434 \\
 & {\em g}$^{\prime}$ & --0.022461 & --0.011719 & 0.012207
 & 0.020020 & --0.018555 & 0.017578 & 0.0466 \\
 & {\em r}$^{\prime}$ & --0.022461 & --0.013184 & 0.013184
 & 0.020020 & --0.019531 & 0.016113 & 0.0441 \\
4734 & {\em u}$^{\prime}$ & --0.018066 & --0.014160 & 0.010742
 & 0.020020 & --0.017090 & 0.017578 & 0.0341 \\
 & {\em g}$^{\prime}$ & --0.022461 & --0.013184 & 0.013672
 & 0.021484 & --0.019531 & 0.017578 & 0.0531 \\
 & {\em r}$^{\prime}$ & --0.022461 & --0.013184 & 0.015137
 & 0.020020 & --0.017090 & 0.017578 & 0.0375 \\
\hline
DV~UMa\\
69023 & {\em u}$^{\prime}$ & --0.033850 & --0.030644 & 0.030276
 & 0.033482 & --0.031445 & 0.032680 & 0.0465 \\
 & {\em g}$^{\prime}$ & --0.033850 & --0.030644 & 0.030276
 & 0.033482 & --0.031445 & 0.032680 & 0.0373 \\
 & {\em i}$^{\prime}$ & --0.033048 & --0.029842 & 0.030276 
 & 0.033482 & --0.030644 & 0.031879 & 0.0245 \\
69046  & {\em u}$^{\prime}$ & --0.032798 & --0.030132 & 0.031210
 & 0.033210 & --0.031465 & 0.031877 & 0.0451 \\
 & {\em g}$^{\prime}$ & --0.033467 & --0.030132 & 0.031210
 & 0.033879 & --0.032132 & 0.032543 & 0.0435 \\
 & {\em i}$^{\prime}$ & --0.032798 & --0.030132 & 0.030544 
 & 0.033210 & --0.030799 & 0.031877 & 0.0239 \\
69058 & {\em u}$^{\prime}$ & --0.032691 & --0.030564 & 0.030612
 & 0.032206 & --0.030564 & 0.032206 & 0.0356 \\
 & {\em g}$^{\prime}$ & --0.033224 & --0.030564 & 0.030612
 & 0.032739 & --0.031097 & 0.032206 & 0.0318 \\
 & {\em i}$^{\prime}$ & --0.033755 & --0.030564 & 0.031142
 & 0.033270 & --0.032161 & 0.032739 & 0.0312 \\
\hline
\end{tabular}
\label{wd_times}
\end{center}
\end{table*}

\begin{table*}
\begin{center}
\caption[]{Bright-spot contact phases.}
\begin{tabular}{cccccccc}
\hline
Cycle & Band & $\phi_{\rmn{b}1}$ & $\phi_{\rmn{b}2}$ &
$\phi_{\rmn{b}3}$ & $\phi_{\rmn{b}4}$ &
$\phi_{\rmn{bi}}$ & $\phi_{\rmn{be}}$\\
\hline
XZ~Eri\\
4733 & {\em u}$^{\prime}$ & --0.000977 & 0.006836 & 0.064941
& 0.067871 & 0.002930 & 0.066406 \\
  & {\em g}$^{\prime}$ & 0.000000 & 0.006836 & 0.069336
& 0.074219 & 0.003906 & 0.073242 \\
  & {\em r}$^{\prime}$ & --0.000977 & 0.006836 & 0.069336
& 0.073242 & 0.001465 & 0.070313 \\
4734 & {\em u}$^{\prime}$ & --0.000977 & 0.006836 & 0.063965
& 0.081055 & 0.002930 & 0.070801 \\
  & {\em g}$^{\prime}$ & --0.000977 & 0.006836 & 0.065430
& 0.070801 & 0.001465 & 0.067871 \\
  & {\em r}$^{\prime}$ & 0.000488 & 0.005859 & 0.065430
& 0.069336 & 0.002930 & 0.067871 \\
\hline
DV~UMa\\
69023 & {\em u}$^{\prime}$ & --0.018620 & --0.009803 & 0.079171
 & 0.083179 & --0.013811 & 0.082378 \\ 
 & {\em g}$^{\prime}$ & --0.016215 & --0.011406 & 0.079171
 & 0.083179 & --0.013811 & 0.080775 \\
 & {\em i}$^{\prime}$ & --0.017017 & --0.009001 & 0.078370
 & 0.085584 & --0.013009 & 0.079973 \\ 
69046 & {\em u}$^{\prime}$ & --0.016797 & --0.012129 & 0.079884
 & 0.085885 & --0.014131 & 0.080553 \\ 
 & {\em g}$^{\prime}$ & --0.018130 & --0.010129 & 0.079884
 & 0.084552 & --0.014131 & 0.081886 \\ 
 & {\em i}$^{\prime}$ & --0.022131 & --0.006127 & 0.081220
 & 0.083219 & --0.014797 & 0.081886 \\ 
69058 & {\em u}$^{\prime}$ & --0.015671 & --0.009819 & 0.079019
 & 0.080613 & --0.014074 & 0.079550 \\ 
 & {\em g}$^{\prime}$ & --0.015671 & --0.010883 & 0.079019
 & 0.081680 & --0.013541 & 0.079550 \\ 
 & {\em i}$^{\prime}$ & --0.014604 & --0.010883 & 0.077955 
 & 0.083274 & --0.013010 & 0.078486 \\ 
\hline

\end{tabular}
\label{bs_times}
\end{center}
\end{table*}

\subsection{A parameterized model of the eclipse}
\label{techniques_model}
Another way of determining the system parameters is to use a physical
model of the binary system to calculate eclipse light curves for each
of the various components. We used the technique developed by
\citet{horne94} and described in detail therein. This model assumes
that the eclipse is caused by the secondary star, which completely
fills its Roche lobe. A few changes were necessary in order to make
the model of \citet{horne94} suitable for our data. The most important
of these was the fitting of the secondary flux, prompted by the
detection of a significant amount of flux from the secondary in the
{\em i}$^{\prime}$ band of DV~UMa. The secondary flux is very small in
all the other bands. Fitting of ellipsoidal variations made no
significant improvement to the overall fit, so we have assumed the
flux from the secondary star to be constant. For both XZ~Eri and
DV~UMa we fitted this model to all the cycles, which were phase-folded
and binned by 2 data points. Examination of the light curves shown in Figs
\ref{light curve} and \ref{dvuma_light curve} shows that
cycle-to-cycle variations for both targets were minimal.

The 10 parameters that control the shape of the light curve are as
follows:
\begin{enumerate}
\item The mass ratio, $q$.
\item The eclipse phase full-width at half-depth, $\Delta\phi$.
\item The outer disc radius, $R_{\rmn{d}}/a$.
\item The white dwarf limb darkening coefficient, $U_{\rmn{w}}$.
\item The white dwarf radius, $R_{\rmn{w}}/a$.
\item The bright-spot scale, $S/a$. The bright-spot is modelled as a
linear strip passing through the intersection of the gas stream and
disc. The intensity distribution along this strip is given by
$(X/S)^{2}e^{-X/S}$, where $X$ is the distance along the strip.
\item The bright-spot tilt angle, $\theta_{\rmn{B}}$,
measured relative to the line joining the white dwarf and the
secondary star. This allows adjustment of the phase of the orbital hump.
\item The fraction of bright-spot light which is isotropic, $f_{iso}$.
\item The disc exponent, $b$, describing the power law of the radial
intensity distribution of the disc.
\item A phase offset, $\phi_{0}$.
\end{enumerate}

The {\sc Amoeba} algorithm (downhill simplex; \citealt{press86}) was
used to adjust selected parameters to find the best fit. A linear
regression was then used to scale the four light curves (for the white
dwarf, bright-spot, accretion disc and secondary) to fit the observed
light curves in each passband. The data were not good enough to
determine the limb-darkening coefficient $U_{\rmn{w}}$ accurately, so
this was held at a typical value of 0.5 for each fit. The disc
parameter for DV~UMa was held fixed at $b=1.0$ as it was too faint to
be well constrained.

In order to estimate the errors on each parameter once the best fit
had been found, we perturbed one parameter from its best fit value by
an arbitrary amount (initially 5 per~cent) and re-optimised the rest of
them (holding the parameter of interest, and any others originally
kept constant, fixed). We then used a bisection method to determine
the perturbation necessary to increase $\chi^{2}$ by $1$, i.e.\
$\chi^{2}-\chi_{\rmn{min}}^{2}=\Delta\chi^{2}=1$. The difference
between the perturbed and best-fit values of the parameter gave the
relevant $1\sigma$ error \citep*{lampton76}. This procedure failed to
find the likely error for the disc exponent $b$ of the {\em
u}$^{\prime}$ band of XZ~Eri, as the disc flux is small in this case
and the light curve noisy, so perturbation of the parameter made
virtually no difference to the $\chi^{2}$ of the fit.

\begin{figure*}
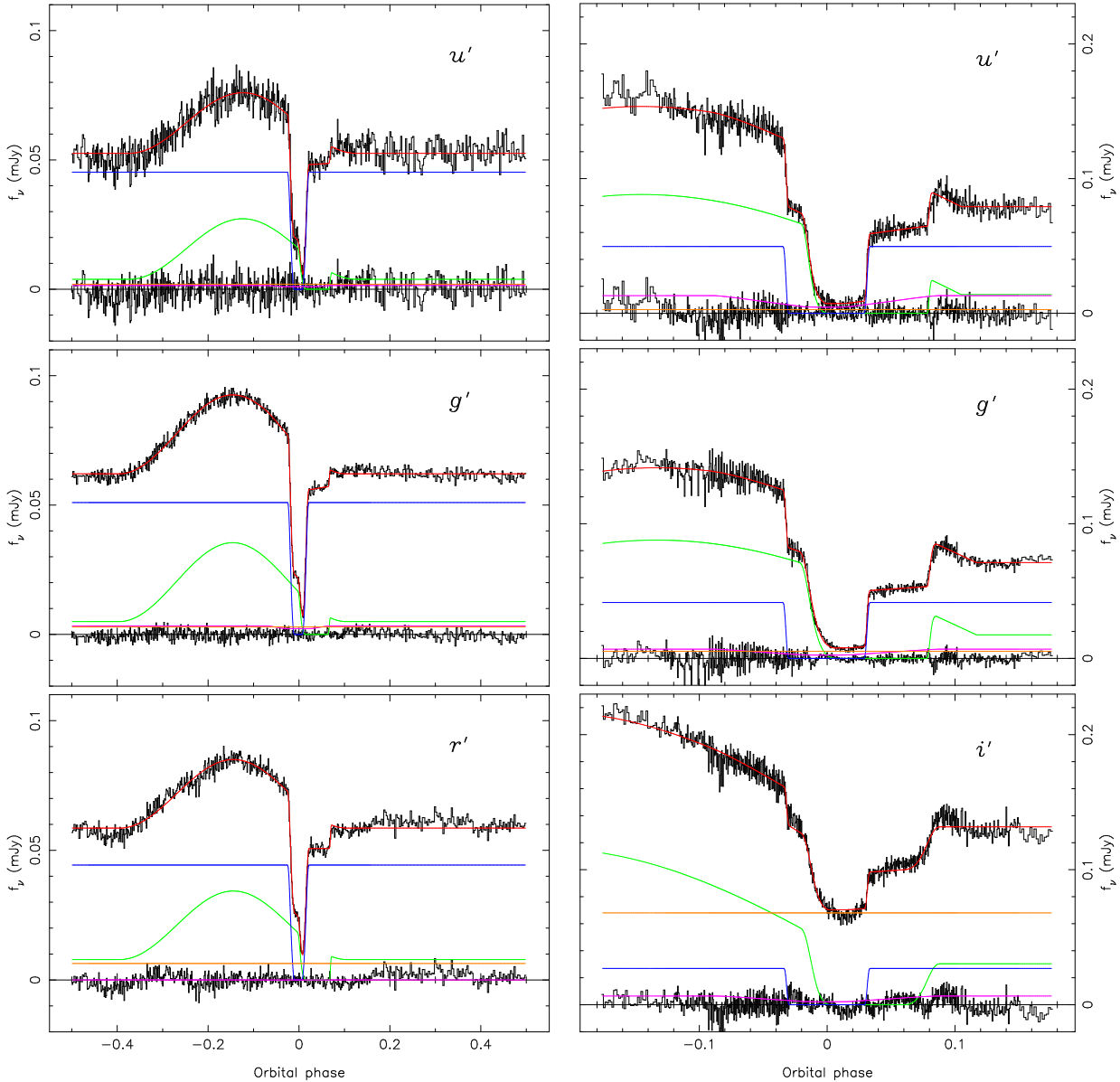

\begin{tabular}{cc}
\psfig{figure=figure4a.ps,width=8.0cm,angle=-90.} &
\psfig{figure=figure4b.ps,width=7.9cm,angle=-90.} \\
\psfig{figure=figure4c.ps,width=8.0cm,angle=-90.} &
\psfig{figure=figure4d.ps,width=7.9cm,angle=-90.} \\
\psfig{figure=figure4e.ps,width=8.0cm,angle=-90.} &
\psfig{figure=figure4f.ps,width=7.9cm,angle=-90.}
\end{tabular}
\caption{Left: the phase-folded {\em u}$^{\prime}$, {\em g}$^{\prime}$
  and {\em r}$^{\prime}$ light curves of XZ~Eri, fitted separately
  using the model described in Section~\ref{techniques_model}. Right:
  the phase-folded {\em u}$^{\prime}$, {\em g}$^{\prime}$ and {\em
  i}$^{\prime}$ light curves of DV~UMa. The data (black) are shown
  with the fit (red) overlaid and the residuals plotted below
  (black). Below are the separate light curves of the white dwarf
  (blue), bright  spot (green), accretion disc (purple) and the
  secondary star (orange). Note that the disc in both objects is very
  faint, as is the secondary (except for the {\em i}$^{\prime}$ band
  of DV~UMa).}
\label{lfit}
\end{figure*}

The results of the model fitting are given in
Table~\ref{parameters_lfit}  and shown in Fig.~\ref{lfit}. Each colour
band was fitted independently, as there were found to be significant
differences between many of the optimum parameters for each band. This
is to be expected for parameters such as the bright-spot scale $S$,
where one would anticipate that the cooler regions are more extended
than the hotter ones (as seen for DV~UMa). We would of course expect
the mass ratio to remain constant in all three colour bands for each
object, which it indeed does.

The results of a white dwarf model atmosphere fit \citep{bergeron95}
to the fluxes in each passband are given in Table~\ref{parameters}.
We have used the white dwarf cooling curves of \citet{wood95} for
$M_{\rmn{w}}/M_{\sun}=0.75$ (interpolating between 0.7 and 0.8) and
$M_{\rmn{w}}/M_{\sun}=1.0$, the approximate masses found using the
Nauenberg relation for XZ~Eri and DV~UMa, to give radial corrections
of 7.6 and 7.0 per~cent, respectively. These were used to determine
the absolute system parameters given in Table~\ref{parameters}.

We note that the higher signal-to-noise light curves of the {\em
i}$^{\prime}$, {\em r}$^{\prime}$ and {\em g}$^{\prime}$ bands have
$\chi^{2}/\nu\gg1$ (see Table~\ref{parameters_lfit}). This is
because these data are dominated by flickering, not photon noise,
unlike the {\em u}$^{\prime}$ data. If we had enough cycles to
completely remove the effects of flickering we would expect, for an
accurate model, to achieve $\chi^{2}/\nu=1$.

Since not all of the orbital cycle of DV~UMa was observed, the
parameters that are constrained by the orbital hump are rather more
uncertain for DV~UMa than they are for XZ~Eri. This may introduce some
systematic errors into the estimation of the bright-spot orientation
$\theta_{\rmn{B}}$ and the bright-spot flux. We suspect that the
slightly unsatisfactory fit to the {\em i}$^{\prime}$ data of DV~UMa
during bright-spot egress is due to additional structure superimposed
on the orbital hump, causing the bright-spot orientation to be
overestimated.

\begin{table}
\begin{center}
\caption[]{Mean position and extent of the bright-spot as
  defined in \citet{feline04a}.}
\begin{tabular}{ccc}
\hline
 & XZ~Eri & DV~UMa\\
\hline
$\Delta R_{\rmn{d}}/a$ & 0.0378 & 0.0258 \\
$\Delta \theta$ & $8\fdg73$ & $7\fdg57$ \\
$\Delta Z/a$ & 0.0174 & 0.0399 \\
$\Delta Z_{2}/a$ & 0.0161 & 0.0217 \\
$R_{\rmn{d}}/a$ & 0.300 & 0.322 \\
$\theta$ & $34\fdg53$ & $27\fdg47 $\\
\hline
\end{tabular}
\label{bs}
\end{center}
\end{table}

\begin{table*}
\begin{center}
\caption[]{Parameters fitted using a modified version of the model of
\citet{horne94}. The fluxes of each component are also
shown. XZ~Eri has been fitted by phase-folding the two eclipses and
binning by two data points. DV~UMa has been fitted by phase-folding all
three eclipses and binning by two data points. Note that the orbital
inclination $i$ is not a fit parameter but is calculated using $q$ and
$\Delta\phi$.}
\begin{tabular}{lcccccc}
\hline
Parameter & \multicolumn{3}{c}{XZ~Eri} & \multicolumn{3}{c}{DV~UMa} \\
 & {\em u}$^{\prime}$ & {\em g}$^{\prime}$ & {\em r}$^{\prime}$ & {\em
 u}$^{\prime}$ & {\em g}$^{\prime}$ & {\em i}$^{\prime}$ \\
\hline
Inclination $i$ & $80\fdg4\pm0\fdg8$ & $80\fdg1\pm0\fdg1$ &
$80\fdg4\pm0\fdg2$ & $83\fdg8\pm0\fdg2$ & $84\fdg3\pm0\fdg1$ &
$84\fdg3\pm0\fdg1$ \\ 
Mass ratio $q$ & $0.11\pm0.02$ & $0.116\pm0.003$ & $0.107\pm0.002$
& $0.159\pm0.003$ & $0.1488\pm0.0011$ & $0.153\pm0.002$ \\ 
Eclipse phase & $0.0342$ & $0.03362$ & $0.0333$ & $0.06346$ & $0.06352$
& $0.06307$\\
\hspace{0.1cm} width $\Delta\phi$ & $\pm0.0007$ & $\pm0.00021$
& $\pm0.0003$ & $\pm 0.00017$ & $\pm0.00007$ & $\pm0.00015$ \\
Outer disc \\
\hspace{0.1cm} radius $R_{\rmn{d}}/a$ & $0.307\pm0.011$ & $0.295\pm0.003$
& $0.316\pm0.005$ & $0.317\pm0.004$ & $0.32278\pm0.00016$ &
$0.31272\pm0.00017$\\
White dwarf \\
\hspace{0.1cm} limb & 0.5 & 0.5 & 0.5 & 0.5 & 0.5 & 0.5 \\
\hspace{0.1cm} darkening $U_{\rmn{w}}$ \\
White dwarf \\
\hspace{0.1cm} radius $R_{\rmn{w}}/a$ & $0.019\pm0.002$ &
$0.0175\pm0.0006$ & $0.0195\pm0.0010$ & $0.0091\pm0.0016$ &
$0.0092\pm0.0004$ & $0.0082\pm0.0014$ \\
Bright-spot \\
\hspace{0.1cm} scale $S/a$ & $0.014\pm0.010$ & $0.013\pm0.002$ &
$0.0147\pm0.0008$ & $0.0150\pm0.0010$ & $0.0211\pm0.0002$ &
$0.049\pm0.003$ \\
Bright-spot \\
\hspace{0.1cm} orientation $\theta_{\rmn{B}}$ & $134\fdg1\pm1\fdg0$ &
$141\fdg9\pm0\fdg3$ & $141\fdg4\pm0\fdg3$ & $142\fdg0\pm0\fdg8$ &
$137\fdg75\pm0\fdg09$ & $169\fdg4\pm0\fdg6$\\ 
Isotropic flux \\
\hspace{0.1cm} fraction $f_{iso}$ & $0.14\pm0.03$ & $0.140\pm0.008$ &
$0.2294\pm0.0015$ & $0.157\pm0.009$ & $0.1989\pm0.0019$ &
$0.262\pm0.004$\\  Disc exponent $b$ & 0.74965 & $0.4\pm2.1$ &
$0.3\pm0.3$ & 1. & 1. & 1.\\ Phase offset $\phi_{0}$ &
$16\times10^{-4}$ & $16.3\times10^{-4}$ & $17.0\times10^{-4}$ &
$2.5\times10^{-4}$ & $5.48\times10^{-4}$ & $1.7\times10^{-4}$\\ &
$\pm3\times10^{-4}$ & $\pm0.8\times10^{-4}$ & $\pm1.2\times10^{-4}$ &
$\pm0.9\times10^{-4}$ & $\pm0.10\times10^{-4}$ &
$\pm0.7\times10^{-4}$\\ 
$\chi^{2}$ of fit & 656 & 897 & 1554 & 1059 & 6873 & 4332\\ 
Number of \\
\hspace{0.1cm} datapoints $\nu$ & 611 & 611 & 611 & 636 & 636 & 636\\
\hline
Flux (mJy) \\
\hspace{0.1cm} White dwarf & $0.0453\pm0.0011$ & $0.0510\pm0.0004$ &
$0.0443\pm0.0004$ & $0.0496\pm0.0008$ & $0.0415\pm0.0002$ & $0.0269\pm0.0004$\\
\hspace{0.1cm} Accretion disc & $0.001\pm0.003$ & $0.0033\pm0.0009$ &
$0.0000\pm0.0010$ & $0.0131\pm0.0015$ & $0.0069\pm0.0004$ & $0.0065\pm0.0007$\\
\hspace{0.1cm} Secondary & $0.0020\pm0.0019$ & $0.0029\pm0.0006$ &
$0.0064\pm0.0007$ &
$0.0027\pm0.0007$ & $0.00531\pm0.00018$ & $0.0680\pm0.0003$\\
\hspace{0.1cm} Bright-spot & $0.0273\pm0.0005$ & $0.03545\pm0.00018$ &
$0.0343\pm0.0002$ & $0.0882\pm0.0005$ & $0.0879\pm0.00014$ &
$0.1157\pm0.0004$\\ 
\hline
\end{tabular}
\label{parameters_lfit}
\end{center}
\end{table*}

\subsection{Comparison of methods}
We have determined the system parameters of the eclipsing dwarf novae
XZ~Eri and DV~UMa through two methods: the derivative method of
\citet{wood86b} and the parameterized model technique of
\citet{horne94}. We proceed to compare these two techniques, first
noting that the system parameters determined by each (given in
Table~\ref{parameters}) are reassuringly in good agreement for the
most part.

Given data with an excellent signal-to-noise ratio (S/N) and covering
many phase-folded cycles, the measurement of the contact phases from
the light-curve derivative is capable of producing accurate and
reliable results \citep[e.g.][]{wood89a}. It is less dependable with
only a few cycles, however, even if they are individually of high
S/N. This is due to flickering having the effect of partially masking
the exact location of the contact phases
$\phi_{1},\ldots,\phi_{4}$. This problem will affect the values for
the deconvolved fluxes of each component and the constraints on the
size of the white dwarf and bright-spot, which are used to determine
the individual component masses. The mid-points of ingress and egress,
especially those of the white dwarf, are generally still well
determined though, since the signal (a peak in the derivative of the
light curve) is large due to the rapid ingress and egress of the
eclipsed body. This makes the determination of the mass ratio and the
orbital inclination relatively simple and reliable. It also means that
this technique is well suited to determining the times of mid-eclipse
in order to calculate the ephemeris.

We believe that the differences between the component masses and radii
of XZ~Eri determined by each technique (Table~\ref{parameters}) are due
to the above effect of flickering. The mass ratios quoted are
consistent with each other, but the relative white dwarf radius
estimated from the derivative method is somewhat smaller than that
determined from the parameterized model ($R_{\rmn{w}}/a=0.012\pm0.002$
and $R_{\rmn{w}}/a=0.0181\pm0.0005$, respectively). This
also affects the estimates of the component radii and masses.

For the purpose of determining the system parameters we
prefer the parameterized model technique over the derivative
method. This is because the former constrains the parameters using all
the points in the light curve to minimise $\chi^{2}$. This procedure
has several advantages:
\begin{enumerate}
\item The value of $\chi^{2}$ provides a reliable estimate of the
  goodness of fit which is used to optimise the parameter
  estimates. The measurement of the contact phases and subsequent
  deconvolution of the light curves in the derivative method is not
  unique (it is affected by the choice of box-car and median filters,
  for instance), and this technique lacks a comparable merit function.
\item Rapid flickering and photon noise during the ingress and/or
  egress phases are less problematic for the parameterized model as the
  light curves are evaluated using all the data points, not just the
  few during ingress and egress.
\item The above points indicate that the parameterized model technique
  requires fewer cycles to obtain accurate results. This is indeed
  what we found in practice, meaning that this method could be applied
  to each passband separately to investigate the temperature
  dependence of each parameter, if any.
\item The bright-spot egress in particular is often faint (due to
  foreshortening) and difficult to reconstruct using the derivative
  method. The parameterized model method is also likely to be easier to
  apply to cases where the ingress of the white dwarf and bright-spot
  are merged, as seen in IP~Peg \citep{wood86b} and EX~Dra
  \citep{baptista00}.
\end{enumerate}

For these reasons, we believe that the results given by the
parameterized model of the eclipse are better determined than those of
the derivative technique. However, the former method does have some
disadvantages. Ideally, it requires observations of most of the
orbital cycle, as the orbital hump is needed to fit some parameters 
reliably. Longer time-scale flickering can also cause some problems
if only a few cycles are available. As with any such technique, the
key weakness of the parameterized model method is the need for an
accurate model. As Fig.~\ref{lfit} shows, apart from the {\em
i}$^{\prime}$ band of DV~UMa the residual from the fit shows no large
peaks in areas such as the ingress and egress of the white dwarf or
bright-spot. Such peaks would be expected if the model were not
adequately fitting the data.

\section{Discussion}
\label{discussion}

We have presented an analysis of two quiescent eclipses of XZ~Eri and three
quiescent eclipses of DV~UMa. For both objects, separate eclipses of
the white dwarf and bright-spot were observed. The identification of
the bright-spot ingress and egress is unambiguous in each case. These
eclipses have been used to determine the system parameters, given in
Table~\ref{parameters}, via two independent methods. The first of
these is through analysis of the light-curve derivative
\citep{wood85,wood86b} and the second by fitting a parameterized model
of the eclipse \citep{horne94}.

The system parameters of DV~UMa have also been estimated by
\citet{patterson00} using eclipse deconvolution. Our analysis is
consistent with their findings, but the parameterized model of the
eclipse provides much more accurate results. The value we obtain for
the mass ratio, for instance, is a factor of $\sim17$ more accurate
than that obtained by \citet{patterson00}. As \citet{patterson00}
note, the spectral type of the secondary star in DV~UMa (M4.5;
\citealp{mukai90}) implies $M_{\rmn{r}}/M_{\sun}=0.12-0.18$ for a
main-sequence star of solar metallicity \citep{chabrier97,henry99},
consistent with our results (Table~\ref{parameters}). \citet{mukai90}
derive the primary temperature and radius of DV~UMa from spectroscopic
observations by assuming that the white dwarf emits a blackbody
spectrum. The temperature they derive, $T_{\rmn{w}}=22000\pm1500$~K,
is consistent with our results (Table \ref{parameters}). The primary
radius ($R_{\rmn{w}}=26000-7700$~km) \citet{mukai90} calculate is only
marginally consistent with our results for the derivative technique
and not consistent with the results of the parameterized model. This
is probably due to the limitation of assuming a blackbody spectrum
\citep{mukai90}.

The two quiescent eclipses of XZ~Eri have been used to make the first
determination of the system parameters for this object, given in
Table~\ref{parameters}. The mass ratio we derive, $q=0.1098\pm0.0017$,
is consistent with XZ~Eri being an SU~UMa star
\citep{whitehurst88,whitehurst91}, as indicated by its (super)outburst
history \citep{woudt01,uemura04}. We also note that the orbital period
and mass ratio of XZ~Eri are similar to those of OY~Car \citep{wood89a}.

\begin{table*}
\begin{center}
\caption[]{System parameters of XZ~Eri and DV~UMa derived using the
  Nauenberg mass--radius relation corrected to the appropriate 
  $T_{\rmn{w}}$. $R_{\rmn{r}}$ is the volume radius of the secondary's
  Roche lobe \citep{eggleton83}, and $R_{\rmn{min}}$ is as defined by
  \citet[][ equation 13]{verbunt88}. The weighted means of the
  appropriate values from Table~\ref{parameters_lfit} are used for the
  model parameters. Each object has one column of parameters
  calculated using the derivative method, and one column derived using
  the parameterized model technique.}
\begin{tabular}{lcccc}
\hline
 & \multicolumn{2}{c}{XZ~Eri} & \multicolumn{2}{c}{DV~UMa} \\
Parameter & Derivative & Model & Derivative & Model \\
\hline
Inclination $i$ & $80\fdg3 \pm 0\fdg6$ & $80\fdg16 \pm 0\fdg09$ &
 $84\fdg4 \pm 0\fdg8$& $84\fdg24\pm0\fdg07$ \\
Mass ratio $q=M_{\rmn{r}}/M_{\rmn{w}}$ & $0.117 \pm 0.015$ & $0.1098
 \pm 0.0017$ & $0.148 \pm 0.013$ & $0.1506\pm0.0009$ \\
White dwarf mass $M_{\rmn{w}}/M_{\sun}$ & $1.01 \pm 0.09$ & $0.767 \pm
 0.018$ & $1.14 \pm 0.12$ & $1.041 \pm 0.024$ \\
Secondary mass $M_{\rmn{r}}/M_{\sun}$ & $0.119 \pm 0.019$ & $0.0842 \pm
 0.0024$ & $0.169\pm0.023$ & $0.157\pm0.004$ \\
White dwarf radius $R_{\rmn{w}}/R_{\sun}$ & $0.0082 \pm 0.0014$ &
 $0.0112 \pm 0.0003$ & $0.0067\pm0.0018$ & $0.0079\pm0.0004$ \\
Secondary radius $R_{\rmn{r}}/R_{\sun}$ & $0.147 \pm 0.015$ & $0.1315
 \pm 0.0019$ & $0.207\pm0.016$ & $0.2022\pm0.0018$ \\
Separation $a/R_{\sun}$ & $0.680 \pm 0.021$ & $0.619 \pm 0.005$ &
 $0.90\pm0.03$ & $0.869\pm0.007$ \\
White dwarf radial velocity $K_{\rmn{w}}/\rmn{km\;s^{-1}}$ & $58 \pm
 8$ & $49.9 \pm 0.9$ & $68\pm6$ & $66.7\pm0.7$ \\
Secondary radial velocity $K_{\rmn{r}}/\rmn{km\;s^{-1}}$ & $496.9 \pm
 2.0$ & $454.7 \pm 0.4$ & $457.5\pm2.6$ & $443.2\pm0.5$ \\
Outer disc radius $R_{\rmn{d}}/a$ & $0.300 \pm 0.017$ & $0.3009 \pm
 0.0025$ & $0.322\pm0.011$ & $0.31805\pm0.00012$ \\
Minimum circularisation radius $R_{\rmn{min}}/a$ & $0.217\pm0.013$ &
 $0.2229\pm0.0014$ & $0.196\pm0.008$ & $0.1948\pm0.0005$ \\
White dwarf temperature $T_{\rmn{w}}/\rm{K}$ & $15000 \pm 500$ &
 $17000 \pm 500$ & $20000\pm1500$ & $20000\pm1500$ \\
\hline
\end{tabular}
\label{parameters}
\end{center}
\end{table*}

The bright-spot scale $S$ of XZ~Eri is constant over all three colour
bands. In DV~UMa, however, it increases in size as the colour becomes
redder. This is easily interpretable: the material cools as it moves
farther from the impact region between the accretion disc and the gas
stream.

\begin{figure}
\centerline{\psfig{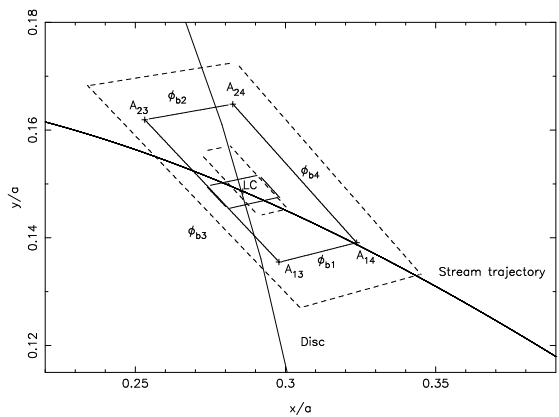}}
\caption{Horizontal structure of the bright-spot of DV~UMa for
  $q=0.148$, showing the region on the orbital plane within which the
  bright-spot lies. The light centre LC is marked by a cross,
  surrounded by the inner solid box, which corresponds to the rms
  variations in position. The phase arcs which correspond to the
  bright-spot contact phases are shown as the outer solid box, with
  the rms variations in position shown as the two dashed
  boxes. Intersections of the phase arcs $\phi_{\rmn{b}j}$ and
  $\phi_{\rmn{b}k}$ are marked $A_{jk}$, with crosses. The stream
  trajectory and disc of radius $R_{\rmn{d}}=0.322a$ are also plotted
  as solid curves.}
\label{dvuma_bs_h}
\end{figure}

The results from the parametrised model of XZ~Eri give a very low
secondary star mass of $M_{\rmn{r}}/M_{\sun}=0.0842\pm0.0024$. This is
close to the upper limit on the mass of a brown dwarf, which is
$0.072M_{\sun}$ for objects with solar composition, but can be up to
$0.086M_{\sun}$ for objects with zero metallicity \citep{basri00}.

The empirical mass-radius and mass-period relations for the secondary
stars of CVs of \citet{smith98a} are in good agreement with the values
determined here. The mass of the white dwarf in XZ~Eri is consistent
with the mean mass of white dwarfs in dwarf novae below the period gap
derived by \citet{smith98a}. The white dwarf in DV~UMa, however, is
unusually massive. Our assumption that we are observing a bare white
dwarf and not a boundary layer around the primary cannot explain this,
as the white dwarf mass derived would be in this case a lower limit
\citep[e.g.][]{feline04a}.

\citet{bisikalo98} found from numerical simulations that `bright-spot'
eclipse features in CVs may be due to an extended shock wave located
on the edge of the stream. Our results do not show any evidence for
this. If the bright-spot emission were coming from a region of shocked
gas in the stream then we might expect the bright-spot orientation
$\theta_{\rmn{B}}$ to coincide with the flow direction of the stream,
which is approximately $169\degr$ for XZ~Eri and $167\degr$ for
DV~UMa. In fact, apart from the less reliable {\em i}$^{\prime}$ band
measurement of DV~UMa, the results in Table~\ref{parameters_lfit} show
that the orientation is half-way between the direction of the stream
and disc (approximately $125\degr$ for XZ~Eri and $118\degr$ for
DV~UMa) flows. The eclipse timings of the bright-spot also show the
bright-spot to be extended along the line between the stream and disc
trajectories (Fig.~\ref{dvuma_bs_h}).

Finally, we note that the system parameters we derive for DV~UMa are
consistent with the superhump period-mass ratio relation of
\citet{patterson98}. XZ~Eri, however, lies $5\sigma$ off this
relation. We use here the superhump periods
$P_{\rmn{sh}}=0.062808\pm0.000017$~days for XZ~Eri \citep{uemura04}
and $P_{\rmn{sh}}=0.08870\pm0.00008$~days for DV~UMa
\citep{patterson00}.

\section*{acknowledgments}
We would like to thank Alan Fitzsimmons for giving us some of his
valuable telescope time. WJF and CSB are supported by PPARC
studentships. TRM acknowledges the support of a PPARC Senior Research
Fellowship. ULTRACAM is supported by PPARC grant
PPA/G/S/2002/00092. This research has made use of NASA's Astrophysics
Data System Bibliographic Services. Based on observations made with
the William Herschel Telescope operated on the island of La Palma by
the Isaac Newton Group in the Spanish Observatorio del Roque de los
Muchachos of the Instituto de Astrofisica de Canarias.

\bibliographystyle{mn2e} \bibliography{abbrev,refs}

\end{document}